
\input phyzzx
\Pubnum={\vbox{ \hbox{CERN-TH.6761/92}}}
\pubnum={CERN-TH.6761/92}
\date={December, 1992}
\pubtype={}
\titlepage

\title{NONPERTURBATIVE SOLUTION OF THE \break
 SUPER-VIRASORO CONSTRAINTS}
\vskip 1.0cm
\author{K. Becker\foot{$^{\dagger}$Permanent
address: Universit\"at Bonn, Physikalisches Institut, Nussallee 12,
W-5300 Bonn 1, Germany.} \break and \break  M. Becker$^{\dagger}$}
\address{Theory Division, CERN\break
 CH-1211 Geneva 23, Switzerland}

\abstract{We present the solution of the discrete
super-Virasoro constraints to all orders of the genus
expansion. Integrating over the fermionic variables we get a
representation of the partition function in terms of the one-matrix
model. We also obtain the nonperturbative solution of the super-Virasoro
constraints in the double scaling limit but  do not find agreement
between our flows and the known supersymmetric extensions of KdV.}

\endpage

\pagenumber=1

\def\rbpz{A. A. Belavin, A. M. Polyakov and A. B. Zamolodchikov,
{\it Nucl.
Phys.} {\bf B 241} (1984), 333}
\def\rberkleb{M. Bershadsky and I. R. Klebanov, {\it Nucl. Phys.}
{\bf B
360} (1991) 559}
\def\rkutasov{ D. Kutasov, ``Irreversibility of the renormalization
group flow in two dimensional quantum gravity'',  Princeton preprint
(1992),
PUPT-1334, hepth@xxx/9207064}

\def\rgl{M. Goulian and M. Li,
{\it Phys. Rev. Lett. } {\bf 66} (1991) 2051;
V. S. Dotsenko, {\it Mod. Phys. Lett.} {bf A6} (1991) 3601;
Y. Kitazawa, {\it Phys. Lett. } {\bf B 265} (1991) 262;
P. Di Francesco and D. Kutasov, {\it Phys. Lett.}{\bf B261}
(1991) 385.}

\def\rsloops{L. Alvarez-Gaum\'e, H. Itoyama, J. Ma\~nes and A.
Zadra, ``Superloop Equations and Two-Dimensional
Supergravity'', CERN-TH.6329/91. {\it Int. J. Mod. Phys.}, to
appear.}
\def\rfive{L. Alvarez-Gaum\'e, K. Becker, M. Becker, R. Empar\'an and
J. Ma\~nes, ``Double Scaling Limit of the Super-Virasoro Constraints'',
CERN-TH.6687/92, hepth@xx x/9207096}
\def\ritoyama{H. Itoyama, ``Integrable Superhierarchy of Discretized 2d
Supergravity'', Stony Brook Preprint ITP-SB-92-21, hepth@xxx/9206091}

\def\rpz{ A.M. Polyakov and
A.B. Zamolodchikov, {\it Mod Phys. Lett.}
{\bf A3} (1988) 1213.}
\def\rdhk{J. Distler, Z. Hlousek and H. Kawai, {\it Int. J. Mod. Phys.}
 {\bf A5} (1990) 391.} \def\rdoubles{ E. Br\'ezin and V.A. Kazakov,
{\it Phys. Lett.} {\bf 236B} (1990) 144;
 M.R. Douglas and S.H. Shenker,
{\it Nucl. Phys.} {\bf B335} (1990) 635;
 D.J. Gross and A.A. Migdal,
{\it Phys. Rev. Lett.} {\bf 64} (1990) 127.
D.J. Gross and A. A. Migdal,
{\it Nucl Phys.} {\bf B340} (1990) 333.}
\def\rdouglas{ M.R. Douglas,
{\it Phys. Lett.} {\bf 238B} (1990) 176.}
\def\rdifdk{ P. Di Francesco,
J. Distler and D. Kutasov,
{\it Mod. Phys. Lett.}{\bf A5} (1990) 2135.}

\def\rdvv{ R. Dijkgraaf, E. Verlinde and
H. Verlinde, {\it Nucl. Phys.} {\bf B348} (1991) 435, M. Fukuma,
H. Kawai and R. Nakayama,{\it Int. J. Mod Phys.} {\bf A6} (1991) 1385.}

\def\rkazakov{ V.A. Kazakov, {\it Mod.
Phys. Lett.} {\bf A4} (1989) 2125.}
\def\rsliouv{ E. Abdalla, M.C.B. Abdalla,
D. Dalmazi and K. Harada, IFT preprint
 IFT-91-0351;
 K. Aoki and E. D'Hoker, UCLA preprint UCLA-91-TEP-33;
 L. Alvarez-Gaum\'e and P. Zaugg, {\it Phys. Lett.} {\bf B273} (1991)
 81.}
\def\rwitkon{ E. Witten,``Two Dimensional
Gravity and Intersection Theory on Moduli
 Space", {\it Surveys in Diff. Geom.}
{\bf 1} (1991) 243, and references therein,
 R. Dijkgraaf and E. Witten, {\it Nucl. Phys.}
{\bf B342} (1990) 486;
 M. Kontsevich, ``Intersection Theory
on the Moduli Space of Curves and the
 Matrix Airy Function", Bonn preprint MPI/91-47;
E. Witten, ``On the Kontsevich Model
and Other Models of Two Dimensional
 Gravity", IAS preprint, HEP-91/24.}
\def\rrabin{ J.M. Rabin, {\it Commun.
Math. Phys.} {\bf 137} (1991) 533,
M. Mulase, `` A new super KP system and a characterization
of the Jacobians of arbitrary algebraic curves'', {\it J. Diff. Geom. },
to appear, B.A. Kuperschmidt, {\it Phys. Lett.}{\bf 102A}
(1984) 213; M. Chaichan and P.P. Kulish, {\it Phys. Lett.}
{\bf 183B} 169;  P. Mathieu, {\it Phys. Lett.}{\bf 203B}
(1988) 287, {\it Lett. Math. Phys.} {\bf 16} (1988) 277,
{\it J. Math. Phys.}{\bf 29} (1988) 287;
A. Bilal and J.-L. Gervais, {\it Phys. Lett.}
{\bf 211B} (1988) 95. }
 
\def\rmmreview{V. Kazakov, ``Bosonic Strings
and String Field Theories in One-Dimensional
 Target Space", lecture given at
Carg\`ese, France, May 1990, to appear in
``Random Surfaces and Quantum gravity", ed.
by O. Alvarez et.al.;
  L. Alvarez-Gaum\'e,
{\it Helv. Phys. Acta} {\bf 64} (1991) 359, and references therein;
P. Ginsparg, ``Matrix Models of 2d Gravity'', Los Alamos Preprint
LA-UR-9999, hepth@xxx/9112013}

\def\rmatsuo{Y. Matsuo, unpublished.}

\def\rdijg{R. Dijkgraaf, ``Intersection theory, Integrable Hierarchies
and Topological Field Theory'', IAS preprint IASSNS-91/91,
hepth@xxx/9210003.}
\def\ragm{L. Alvarez-Gaum\'e and J.L. Ma\~nes, {\it
Mod. Phys. Lett.} {\bf A6} (1991) 2039.}


\def\D{\Delta}

\def\p{\partial}

\def\L{\Lambda}
\def\l{\lambda}

\def\k{\kappa^2}

\def\l{\lambda}

\def\DDLR{{\buildrel \leftrightarrow \over D}}

\def\ai{$$
{\cal Z_B}(g_k,N)=\int d^{N^2} \Phi exp[-{N\over \L_{\cal B}} tr
\sum_{k \geq 0} g_k \Phi^k] \qquad , \qquad
\L_{\cal B} =e^{-\mu}
\eqno(1)
$$}
\def\aii{$$
L_n {\cal Z_B} =0\qquad, \qquad n\geq -1 \qquad ; \qquad
[L_n,L_m]=(n-m)L_{n+m} $$
$$
L_n={\L_{\cal B}^2\over N^2}\sum_{k=0}^{n}{\partial^2\over\partial
g_{n-k}\partial g_k}+\sum_{k\geq 0}k g_k{\partial \over \partial
g_{k+n}},
\eqno(2)
$$}
\def\aiii{$$
\alpha_{-n}=-{N\over \L_{\cal B} \sqrt{2}}n g_n,\quad n>0 \qquad
; \qquad \alpha_n =-{\L_{\cal B} \sqrt{2} \over N}{\partial \over
\partial g_n}, \quad n\geq 0\;\;.
\eqno(3)
$$}
\def\aiv{$$
\lim_{N\rightarrow \infty}{1\over N}{\cal Z_B}^{-1}\p X^{-} {\cal Z_B}
 \sim
V'(p)=\sum_{k>0} k g_k p^{k-1},
\eqno(4)
$$}

\def\av{$$
[L_m,L_n]=(n-m)L_{n+m}+{(m^3-m)\over 8}\delta_{n+m,0}\;\; ,
\eqno(5)
$$}
\def\avi{$$
[L_m,G_r]=({m\over 2}-r)G_{m+r}\;\; ,
\eqno(6)
$$}
\def\avii{$$
\{ G_r, G_s \}=2L_{r+s}+{1\over 2}(r^2-{1\over 4})\delta_{r+s,0}\;\; ,
\eqno(7)
$$}

\def\aviii{$$
\alpha_p=-{\L_S\over N}{\partial \over \partial
g_p}\qquad;\qquad \alpha_{-p}=-{N\over \L_S}p g_p\quad,\quad
p=0,1,2,\dots
$$
$$
b_{p+{1\over 2}}=-{\L_S \over N}{\partial \over\partial
\xi_{p+{1 \over 2}}} \qquad;\qquad b_{-p-{1\over 2}}=-{N\over
\L_S}\xi_{p+{1\over 2}}\quad,\quad p=0,1,2,\dots\;\; .
\eqno(8)
$$}
\def\aix{$$
G_{n-{1\over 2}}=\sum_{k=0}^\infty \xi_{k+{1\over 2}}{\partial \over
\partial g_{k+n}}+\sum_{k=0}^\infty kg_k{\partial\over\partial
\xi_{k+n-{1\over 2}}}+{\L_S^2\over
N^2}\sum_{k=0}^{n-1}{\partial\over \partial
\xi_{k+{1\over 2}}}{\partial\over \partial g_{n-1-k}} \quad n\geq 0\;\;,
\eqno(9)
$$}
\def\ax{$$
L_n=\sum_{k=1}^\infty kg_k{\p \over \p g_{k+n}}+\sum_{r={1\over 2}}^
\infty
({n\over 2}+r)\xi_r {\p \over \p \xi_{n+r}}+
{\L_S^2\over 2N^2}\sum_{k=0}^{n}{\p^2\over \p g_k \p g_{n-k}}+
$$
$$
{\L_S^2\over 2N^2}\sum_{r={1\over 2}}^{n-{1\over 2}}({n\over 2}+r)
{\p^2\over \p \xi_r \p \xi_{n-r}}\quad n\geq -1 \;\;.
\eqno(10)
$$}

\def\axi{$$
V(p,\Pi)=\sum_{k\geq 0}(g_k p^k+\Pi \xi_{k+1/2}p^k)\;\;.
\eqno(11)
$$}

\def\axii{$$
Z_S(g_k,\xi_{k+{1\over 2}},N)=\int \prod_{i=1}^N d\lambda_i d\theta_i
\Delta(\lambda,\theta) exp\left({-{N\over \L_S}
\sum_{k\geq0}\sum_{i=1}^{N}(g_k\lambda_i^k
+\xi_{k+{1\over 2}}\theta_i\lambda_i^{k})}\right) \;\;.
\eqno(12)
$$}
\def\axiii{$$
\sum_{i}\lambda_i^n(-{\partial\over\partial\theta_i}
+\theta_i{\partial\over\partial\lambda_i})\Delta=\Delta
\sum_{i\neq j} \theta_i{\lambda_i^n-\lambda_j^n\over
\lambda_i-\lambda_j}\;\; ,
\eqno(13)
$$}

\def\axiv{$$
Z_S=\int \prod_{i=1}^{N} d\lambda_i d\theta_i
\prod_{i<j}(\lambda_i-\lambda_j -\theta_i\theta_j) exp\left(
-{N\over \L_S}\sum_{k\geq 0}\sum_{i=1}^N(g_k\l_i^k+\xi_{k+{1\over 2}}
\theta_i \l_i^k) \right) \;\;.
\eqno(14)
$$}

\def\ci{$$
Z_S(g_k,\xi_{k+{1\over
2}}=0,2N)=\int\prod_{i=1}^{2N}d\l_i \D(\l) Pf^{2N}
(\l_{ij}^{-1}){\rm exp} (-{2N\over \L_S}\sum_k\sum_{i=1}^{2N}
g_k\lambda_i^k)  \;\;.
\eqno(15)
$$}

\def\cii{$$
Z_S(g_k,\xi_{k+{1\over 2}}=0,2N)={1\over 2^N} {2N \choose N} {\cal
Z_B^{\rm 2}}(g_k,N) \;\;.
\eqno(16)
$$}
\def\ciii{$$
 Pf^{2N}(\l_{ij}^{-1})\prod_{i<j}^{2N}(\l_i-\l_j) ={1\over 2^{N}}
\sum_{I,J}
S^{I,J}\Delta^2(I)\Delta^2(J) \;\;,
\eqno(17)
$$}
\def\civ{$$
P(\l)=(\l-\l_2)(\l-\l_3)\l_{12}\l_{13}-(\l-\l_1)(\l-\l_3)\l_{12}\l_{23}+
(\l-\l_1)(\l-\l_2)\l_{13}\l_{23}
\eqno(18)
$$}
\def\cv{$$
Q(\l)={1\over 2}[(\l-\l_3)^2\l_{12}^2+(\l-\l_2)^2\l_{13}^2+
(\l-\l_1)^2\l_{23}^2] \;\; .
\eqno(19)
$$}
\def\cvi{$$
Q(\l_{2N+1})=
{1\over 2^{N}}
\sum_{I,J}
S^{I,J}\Delta^2(I)\Delta^2(J)
\prod_{\scriptstyle i_1< \dots <i_{N}\atop \scriptstyle j_1<\dots
<j_{N}}
(\l_i-\l_{2N+1})^2(\l_j-\l_{2N+1})^2
$$
$$=
Pf^{2N}(\l_{ij}^{-1})\prod_{i<j}^{2N}(\l_{i}-\l_j)
\prod_{i_1<\dots<i_N}(\l_i-\l_{2N+1})^2
\prod_{j_1<\dots<j_N}(\l_j-\l_{2N+1})^2=P(\l_{2N+1})\;\; .
\eqno(20)
$$}

\def\cvii{$$
F_S^{(0)}={{\cal F_B}\over 2}\;\;.
\eqno(21)
$$}

\def\cviii{$$
\L_S=2\sum_{k=1}^\infty kg_k\p_{\L_{\cal B}}\left(-\L_{\cal B}^2{\p
 {\cal F_B}\over \p g_k}\right)\;\; .
\eqno(22)
$$}

\def\cix{$$
{\p^2 Z_S(2N) \over \p \xi_{k+{1\over 2}}\p
\xi_{n+{1\over 2}}}={1\over 2^{N-1}}{ 2N \choose N}
\left(
{\p^2 {\cal Z_B}(N)\over \p g_{k+1}\p g_n}{\cal Z_B}(N)-{\p {\cal
Z_B}(N) \over \p g_{k+1}}{\p {\cal Z_B}(N)\over \p g_n}\right)
-(k\leftrightarrow n) \;\; .
\eqno(23)
$$}

\def\cxa{$$
\sum_{\alpha,\beta=1}^{2N} \l_\alpha^k \l_\beta^n \int \prod_{i=1}^{2N}
d\theta_i \D (\l ,\theta) \theta_\alpha \theta_\beta=
$$
$$
{1\over 2^{N-1}}
\sum_{I,J}
S^{I,J}\Delta^2(I)\Delta^2(J)
\sum_{\alpha,\beta=1}^N (\l_{i_{\alpha}}^{k+1}\l_{i_{\beta}}^n
-\l_{i_{\alpha}}^{k+1}\l_{j_{\beta}}^n)-(k\leftrightarrow n) \;\; ,
\eqno(24)
$$}

\def\cxi{$$
F_S={{\cal F_B}\over 2}-{1\over 2}\sum_{k,n}\xi_{k+{1\over
2}}\xi_{n+{1\over 2}}{\p^2 {\cal F_B}\over \p g_{k+1} \p g_n}\;\; .
\eqno(25)
$$}

\def\doi{$$
T_F={1\over 2}\p \varphi(z)\psi(z)\qquad  , \qquad T_B={1\over
2}:\p \varphi(z)\p \varphi(z):+{1\over 2}:\p \psi(z)\psi(z):+
{1\over 8z^2}
\;\; ,
\eqno(27)
$$}

\def\di{$$
G_{n+{1\over 2}}={t_0\tau_0\over 4\k}\delta_{n,-1}+\sum_{k\geq
0}(k+{1\over 2})t_k{\p \over \p \tau_{n+k+1}}+ {\tau_0\over 2}{\p
\over \p t_n}+\sum_{k\geq 0}\tau_{k+1}{\p \over \p t_{n+k+1}}
+\k \sum _{k=0}^{n}{\p^2\over \p t_k \p \tau_{n-k}}
\eqno(28)
$$}

\def\dii{$$
G_{n+{1\over 2}} Z_S(t_k,\tau_k)=0\qquad, \qquad n\geq -1 \;\;.
\eqno(29)
$$}

\def\diii{$$
\k DF_S=D^{-1}u-2\tau D \tau \;\; ,
\eqno(30)
$$}
\def\div{$$
t=-\sum_{k\geq 1}(2k+1) t_k R_k[u] \qquad, \qquad
\tau=-\sum_{k\geq 0}\tau_k R_k[u] \;\; ,
\eqno(31)
$$}

\def\dv{$$
{\p u \over \p t_n}=D R_{n+1}[u]
\eqno(32)
$$}

\def\dvi{
$$
DR_{n+1}[u]=(\k D^3+4uD+2(Du))R_n[u]\;\; ,
\eqno(33)
$$
$R_0={1\over 2}$, $R_1=u$, $R_2=(3u^2+\k D^2 u)$, $\dots$}

\def\dvii{$$
F_S={\cal F_B}-\sum_{{\scriptstyle n\geq 0 \atop \scriptstyle k
\geq 1}}
\tau_n \tau_k
{\p^2{\cal F_B}\over \p t_n\p t_{k-1}}\;\;  .
\eqno(34)
$$}

\def\dviii{$$
{\p  U \over \p \tau_k} =2D (R_k[u] \DDLR \tau )\qquad , \qquad
{\p U \over \p t_k} =DR_{k+1}[u]-2{\p \over \p t_k}(\tau D^2\tau)\;\; .
\eqno(35)
$$}

\def\dx{$$
{\p u \over \p t_1}=D(3 u^2+\k D^2 u)\qquad,\qquad
{\p \tau \over \p t_1}=6uD\tau +\k D^3 \tau \;\;,
\eqno(36)
$$}

\def\dxib{$$
{\p W_0 \over \p t}={1\over 2}\p_x^3W_0+3W_0\p_x W_0 +{3\over 2}
\p_x^2W_1
W_1 \qquad, \qquad
{\p W_1 \over \p t}={1\over 2}\p_x^3 W_1 +{3\over
2}\p_x(W_0 W_1)
\eqno(37)
$$}

\def\dxii{$$
{\p U\over \p t_1}=D(3U^2+\k D^2U-12DU\tau D\tau +6\k D\tau D^3\tau )
\;\; ,
\eqno(38)
$$}
\def\dxiii{$$
{\p \tau \over \p t_1}=6UD\tau -12\tau D\tau D^2\tau +\k D^3\tau\;\; .
\eqno(39)
$$}

\def\dxiv{$$
\pmatrix{ {\p u \over \p t_{k+1}} \cr {\p \tau \over \p t_{k+1}}\cr}=
\pmatrix{ \k D^2+2u+2DuD^{-1} & 0 \cr
2D\tau D^{-1} +2D^{-1} D\tau & \k D^2+2u+2D^{-1}u D \cr }
\pmatrix{ {\p u \over \p t_k} \cr {\p \tau \over \p t_k}\cr }
\eqno(40)
$$}

\def\dxv{$$
{\p u \over \p \tau_k}=0 \qquad, \qquad
{\p \tau \over \p \tau_{k+1}}=(\k D^2+2u+2D^{-1}uD){\p \tau \over \p
\tau_k}
\eqno(41)
$$}

\par{\bf 1.Introduction.}

Matrix models provide us with powerful methods
to evaluate the sum over geometries in two dimensional quantum gravity.
For $c\leq 1$ conformal field
theories \REF\bpz{\rbpz} [\bpz] coupled to gravity, $n$-point
correlation functions of gravitationally dressed primary fields have
been calculated within this formulation on world sheets with arbitrary
topology, a  result
that is far from being obtained from the continuum Liouville approach
\REF\gl{\rgl}[\gl]. It turns out that matrix models have a close
connection
to integrable systems \REF\lagcgl{L. Alvarez-Gaum\'e, C. Gomez
and J. Lacki {\it Phys. Lett. }{\bf B253} (1991) 56; E. Martinec,
{\it Comm.
Math. Phys.} {\bf 138} (1991) 437.}[\lagcgl], which is manifest
through  the
appearence of the Korteweg-de Vries hierarchy in the double scaling
limit
\REF\doubles{\rdoubles}[\doubles] \REF\douglas{\rdouglas}[\douglas].
The
recent discovery of Kontsevich matrix integrals has shown a natural
connection to topological gravity  \REF\witkon{\rwitkon}[\witkon]
\REF\dijg{\rdijg}[\dijg]. Despite of this important results, the
generalization to the supersymmetric case is connected with
difficulties.
So for example, the role which the known supersymmetric extensions of
KdV \REF\rabin{\rrabin}[\rabin] play in this context remains obscure
\REF\berkleb{\rberkleb}\REF\difdk{\rdifdk} [\berkleb][\difdk] .
For $N=1$ superconformal field theories coupled to two dimensional
supergravity correlation functions for spherical topologies have been
obtained in \REF\sliouv{\rsliouv}[\sliouv], using the continuum
super-Liouville approach. The torus path integral and the spectrum of
physical operators for ${\widehat c}\leq 1$ superconformal models
coupled to 2D-supergravity have been considered in
\REF\kutasov{\rkutasov} [\berkleb] [\kutasov]. However
the more powerful counterpart in terms of super-matrix models has not
 yet
been found \REF\agm{\ragm}[\agm]. To shed some light on this situation,
the authors of \REF\sloops{\rsloops}[\sloops] proposed an
``eigenvalue model'',
which is described in terms of $N$ even and $N$-Grassmann odd
variables. The basic idea for the construction of the model was the
generalization of the  Virasoro constraints known from the
ordinary one matrix model \REF\matsuo{\rmatsuo}\REF\dvv{\rdvv}
[\matsuo] [\dvv] to the $N=1$ supersymmetric case. Generalizing
Kazakov's loop equations  \REF\kazakov{\rkazakov}[\kazakov] to $N=1$
superloop equations , the
 model
was evaluated for spherical topologies in [\sloops], obtaining
agreement with correlation functions [\sliouv] and critical exponents
\REF\dhk{\rdhk}\REF\pz{\rpz} [\dhk] [\pz] of $(2,4m)$
superconformal field theories coupled to world-sheet supergravity.
Further
analysis of the model carried out in
\REF\five{\rfive}\REF\itoyama{\ritoyama} [\five] [\itoyama] showed,
that the
continuum super-Virasoro constraints are described in terms of a
 ${\widehat
c}=1$ theory, with a $Z_2$-twisted scalar and a Weyl Majorana fermion
in the Ramond sector. The solution of the continuum superloop
equations  in the  first few orders of the genus expansion and a
general heuristic
argument showed [\five], that there might be a close relation between
the bosonic piece of the free energy and the free energy of the one
matrix model.

In this letter we analyze the solution of this model to all orders of
the
genus expansion. After a brief review about its construction (section
two),
we analyze the solution of the discrete super-Virasoro constraints in
section three. Integrating over the fermionic variables, we get a
representation of the partition function in terms of the one matrix
model.
In section four we analyze the solution of the super-Virasoro
constraints in
the double scaling limit and compare our flows with the known
supersymmetric extensions of KdV [\rabin].

\par{\bf 2.Construction of the model.}

The bosonic one matrix model is defined as:
\ai
where $\Phi$ is a Hermitian $N\times N$ matrix and $\mu$ is the bare
cosmological constant. It satisfies the Virasoro constraints [\matsuo]:
\aii
which can be derived from (1) by implementing invariance under the
shift $\Phi\rightarrow \Phi+\varepsilon \Phi^{n+1}$.

We obtain a relation to a free massless boson, if we introduce an
infinite set of creation and annihilation operators related to the
coupling constants
\aiii
The Virasoro generators $L_n={1\over 2}\sum :\alpha_{-k}\alpha_{k+n}:$
are
the modes of the energy momentum tensor $T(p)=\sum L_n p^{-n-2}$ of the
scalar field $\p X(p)=\sum \alpha_n p^{-n-1}=\p X^{+}(p)+\p X^{-}(p)$.
{}From  (1) and (3) we see that the potential of the model is related to
the negative modes of the scalar field

\aiv
As
a generalization to the $N=1$ supersymmetric case, we take a ${\widehat
c}=1$ free massless superfield $X(p,\Pi)=x(p)+\Pi \psi(p)$ with energy
momentum tensor $T(p,\Pi)=DX\p X=\psi \p_p x+\Pi:(\p_px\p_px+\p \psi
\psi):$ and the mode expansion $\p X(p)=\sum \alpha_n p^{-n-1}$,
$\psi(p)=\sum b_r p^{-r-1/2}$. The relation to the ${\widehat c}=1$
super-Virasoro algebra:

\av
\avi
\avii
and the coupling constants of the model is introduced, if we define the
modes as:

\aviii

In terms of the coupling constants the Virasoro generators then become:

\aix
\ax

In analogy to the one matrix model the potential is determined from the
negative modes of the superfield $DV(p,\Pi)\sim DX^-$:

\axi

This suggests the introduction of $N$ even-and $N$ Grassmann-odd
eigenvalues, obtaining for the partition function:

\axii

The generalization of the usual Vandermonde determinant,
$\Delta(\lambda,\theta)$, is determined by implementing invariance of
(12) under the super-Virasoro constraints (9) (10), which yields a
differential equation for the measure:

\axiii

whose unique solution up to an irrelevant constant is
$\Delta=\prod_{i<j}(\lambda_i-\lambda_j-\theta_i\theta_j)$. The model
that we want to solve in the large $N$ limit is thus:

\axiv

In (14) $N$ is taken to be even in order to get a Grassmann even
partition function.

\par{{\bf 3. Solution of the discrete model.}}

We analyze the solution of the discrete superloop equations [\sloops]
[\five] or equivalently the discrete super-Virasoro constraints
through the representation (14) of the partition function. We first
consider the  case without fermionic couplings $\xi_{k+{1\over 2}}=0$.
The partition function can be written as a function of the Pfaffian of
the antisymmetric matrix $M_{ij}=\l_{ij}^{-1}$, where then
$\l_{ij}=\l_i-\l_j$, $i\neq j$,  and the usual Vandermonde
$\D(\l)=\prod_{i<j}(\l_i-\l_j)$ [\sloops]:

\ci

Here the product of the Vandermonde and the Pfaffian is totally
symmetric under exchange of two eigenvalues. We want to show, that the
relation of the partition function (15) to the bosonic
partition function of the one matrix model (1) is given
by:

\cii

To prove (16) we show, that the following equivalence for the measures
holds:

\ciii
Here we have introduced the notation $I=(\l_{i_1},\dots,\l_{i_N})$,
$i_1<\dots<i_N$ (same for J), and
$S^{I,J}$ means symmetric permutations between the elements of $I$ and
$J$. The equality (16) is a consequence of (17), since the partition
function (15) with $2N$ eigenvalues will then factorize into
$ 2N \choose N$ products of two independent one matrix models with $N$
eigenvalues.

Obviously (16) holds for the case of two eigenvalues, so that to
illustrate the situation we start with the first nontrivial case of
four eigenvalues, $N=2$. We will consider the right and left hand  side
of  (17) as polynomials $Q(\l)$, $P(\l)$ respectively in $\l=\l_4$:

\civ
\cv
{}From $P(\l_i)=Q(\l_i)$ for $i=1,2,3$ it follows $P(\l)=Q(\l)$, $\forall
\l$.

We show that (17) is correct by induction over $N$.
For $N>2$ we set $\lambda=\lambda_{2N+2}$ so that $P(\l)$, $Q(\l)$
become polynomials of degree $2N-2$ in $\lambda$. It is
enough to show that they coincide for $\lambda=\lambda_{2N+1}$
since the symmetry under the interchange of $\l_i$ and $\l_j$
guarantees $P(\l_i)=Q(\l_j)$ for $i=1,\dots,2N$ and thus $Q(\l)=P(\l)$
$\forall N$. If we consider both polynomials of (17) for $N\rightarrow
N+1$ and $\l=\l_{2N+1}$ we obtain the desidered result by applying the
induction hypothesis.

\cvi

We conclude that the bosonic part of
the supersymmetric $(2N\times 2N)$ model is proportional to the square
of the corresponding $(N\times N)$ bosonic one matrix model (16). With
this result we can already find the form of the discrete string
equation for arbitrary genus.

Writing $Z=e^{N^2 F_S}$, equation (16) implies, that the fermionic
independent piece of the supersymmetric free energy $F_S^{(0)}$ is
related to the free energy of the one matrix model ${\cal F_B}$ (up to
an irrelevant additive constant):

\cvii
Since the dependence of the free energy on $g_0$ is trivial
${\p  F\over \p g_0}=-1/\L$, (21) implies a rescaling of the
cosmological constant $\L_S=2\L_{\cal B}$. This relation can also be
obtained, if we evaluate the bosonic piece of the $L_0$ constraint
(10):

\cviii

Up to a factor of two the cosmological constant satisfies the
same string equation as in the one matrix model. Including the
fermionic couplings, the following equality for the second order in
fermions holds:

\cix

The proof of this relation is straightforward, using the relation:

\cxa

which can be shown using equation (17). Formula (23) is
purely bosonic in the sense, that after derivation we set the
fermionic couplings to zero.

Equations (21) and (23) imply that the free energy until the second
order in fermionic couplings is given by:

\cxi

Since the solution to the continuum super-Virasoro constraints will be
analyzed in a moment, it will then become clear that (25) is the
complete expansion of the free energy and no dependence on more than
two fermionic couplings appears.

Equation (25) gives a representation of the model (12) in terms of
the one matrix model since the fermionic variables $\theta_i$ have been
integrated out. We think that further analysis of this interesting
relation to the one matrix model for the discrete theory could lead to
the geometrical interpretation of the model and the derivation of
Feynman rules. We hope to report on this results elsewhere and turn now
to the solution of the super-Virasoro constraints in the double scaling
limit.

\par{\bf 4.Solution to the super-Virasoro constraints in the double
scaling limit.}

The continuum super-Virasoro constraints [\five][\itoyama] are
described by the super-energy momentum tensor of a ${\widehat c}=1$
superconformal field theory

\doi

where $\varphi(z)$ is a bosonic scalar field with antiperiodic
boundary conditions and $\psi(z)$ is a Weyl-Majorana Fermion in the
Ramond sector. The modes of the fermionic part of the energy momentum
tensor $T_F={1\over 2}\sum G_{n+{1\over 2}}z^{-n-2}$ are described in
terms of the coupling constants of the model for $n\geq -1$:

\di

and the bosonic modes are obtained by the anticommutation relations
(7). The constraints imposed on the partition function
$Z_S(t_k,\tau_k)=exp(F_S)$ are:

\dii

The equations which describe the model (14) in the double scaling
limit are determined by (28) (29):

\diii
\div

where $D={\p \over \p t_0}$, ${\widehat \tau}=\tau+{\tau_0\over 2}
={\p F_S \over \p \tau_0}$ is the first fermionic scaling variable, $t$
is the renormalized cosmological constant and $u$ is the bosonic piece
of the two point function of the puncture operator $\langle
\sigma_0\sigma_0\rangle$, satisfying the KdV-flow equation

\dv

$R_k[u]$ are the  Gel'fand-Dikii polynomials defined through the
recursion relations:

\dvi

To show that (30) (31) is the solution
to (29) it is enough to show that $G_{-{1\over 2}}$ and $G_{3\over 2}$
are satisfied, since the algebra (5)-(7) guarantees for all the
remaining constraints. That this is indeed the case, can be shown by
simple calculations, where we have used well known relations of the one
matrix model theory like (33). The equations (30) (31) coincide with
the solution presented in [\sloops][\five] for the first few orders of
the genus expansion.  The free energy (30) can be written as a function
of the free energy of the one matrix model, which is very similar to
the relation (26) of the discrete model and has an expansion at most
bilinear in the fermionic couplings

\dvii

The form of the solution obtained (30) (31) implies, that the bosonic
piece of the model is equivalent to an ordinary one matrix model, where
the constraints act on the square root of the partition function $L_n
\sqrt{{\cal Z_B}}=0, \;\; n\geq -1$ [\dvv]. The fermionic dependence of
the free energy determines, that correlation functions involving more
than two fermionic operators vanish, since the fermionic scaling
variable ${\widehat \tau}$ is linear in fermionic couplings. These
properties were known to hold for the first orders of the genus
expansion [\sloops][\five] and are in agreement with the previous
analysis of the discrete model.

The correlation functions in the double
scaling limit which follow from (30) (31) as a function of the
two point function of the puncture operator $U=u-2\tau D^2 \tau$ are
given by:

\dviii

The scaling dimensions of the scaling operators (35)
and the string susceptibility (31) are in agreement with
the results of $(2,4m)$-minimal superconformal models coupled to two
dimensional supergravity [\sloops] [\dhk] [\pz].

{}From (35) we obtain for the first nontrivial bosonic flow, a set of
two equations:

\dx

where the first one is the ordinary KdV equation. Equations (36)
are invariant under the global supersymmetric transformations
$\delta u=\epsilon D \tau $, $\delta \tau=\epsilon u $, where
$\epsilon$ is a constant anticommuting parameter, as it is known to
hold for the supersymmetric extensions of KdV of Manin-Radul or
Mulase-Rabin  [\rabin]:

\dxib

Here $W_0$ is a bosonic variable, while $W_1$ is fermionic.
We can write the form of our generalized KdV-equation in terms of
$U$, the perhaps more convenient variable:

\dxii
\dxiii

Equations (38) and (39) can independently be derived
from the first orders of the genus expansion, with the method proposed
in [\five].  From (36) or (38) (39) we have not been able to see any
relation to the known supersymmetric extensions of the KdV-equation
[\rabin], so for example, the perhaps most natural identification
$W_0\sim U$ is not consistent with (37) nor the other extensions of KdV
[\rabin]. We may conclude that our correlation functions (35) describe
a new supersymmetric extension of the KdV-hierarchy.

The bosonic flows satisfy the recursion relations which
can be obtained from the recursion relations of the $R_k[u]$'s:

\dxiv

While for the fermionic flows we have simply:
\dxv

It would be important to study further
properties of the differential equations which describe our correlation
functions to see whether they share the standard properties of
completely super-integrable systems. This will be left for work in the
future.

\vskip 0.5cm

\par{\bf Acknowledgement.}

We would specially like to thank L. Alvarez-Gaum\'e for
many valuable discussions. We have had also interesting discussions
about supersymmetric hierarchies with M. Mulase and J. Rabin. We are
grateful to W. Nahm for his constant help. Finally we thank the Theory
Division of CERN for hospitality.

\REF\rewiev{\rmmreview}


\vfill
\endpage
\refout
\end